\documentstyle[pra,aps,psfig]{revtex}
\newcommand{\beq}{\begin{equation}}
\newcommand{\eeq}{\end{equation}}
\newcommand{\beqa}{\begin{eqnarray}}
\newcommand{\eeqa}{\end{eqnarray}}
\newcommand{\ba}{\begin{array}}
\newcommand{\ea}{\end{array}} 

\begin{document} 
\draft 

\twocolumn[\hsize\textwidth\columnwidth\hsize\csname
@twocolumnfalse\endcsname

\widetext 

\title{Quantum Phases of Attractive Matter Waves in a Toroidal Trap} 
\author{A. Parola$^{1}$, L. Salasnich$^{2}$, R. Rota$^{2}$ 
and L. Reatto$^{2}$} 
\address{$^{1}$Dipartimento di Fisica e Matematica, 
Universit\`a dell'Insubria, \\ 
Via Valleggio 11, 22100 Como, Italy\\
$^{2}$Dipartimento di Fisica and CNR-INFM, Universit\`a di Milano, \\ 
Via Celoria 16, 20133 Milano, Italy} 

\maketitle

\begin{abstract} 
Investigating the quantum phase transition in a ring 
from a uniform attractive Bose-Einstein condensate to a localized 
bright soliton we find that the soliton undergoes transverse collapse 
at a critical interaction strength, which depends on the ring dimensions. 
In addition, we predict the existence 
of other soliton configurations with many peaks, 
showing that they have a limited stability domain. 
Finally, we show that the phase diagram displays several new features 
when the toroidal trap is set in rotation. 
\end{abstract} 


]

\narrowtext 

\section{Introduction}

Metastable states of Bose Einstein condensates (BECs) made 
of attractive $^{7}$Li atoms, 
namely single and multiple bright soliton configurations, 
have been observed in two different experiments \cite{r1,r2}. 
These metastable bright solitons, which can travel 
for long distances without dispersion, 
have been the subject of various theoretical 
investigations because of their relevance in nonlinear 
atom optics \cite{r3,r4}. Recently, repulsive BECs 
in a quasi-1D ring have been produced and studied \cite{r5}.  
The case of an attractive BEC in a ring has not yet been 
experimentally investigated but appears very interesting. 
For this system a quantum phase transition from 
a uniform condensate to a bright soliton. 
has been predicted by G.M. Kavoulakis and by R. Kanamoto, 
H. Saito and M. Ueda \cite{r6}.  
This prediction is based on mean-field and beyond mean-field 
numerical results for a 1D Bose gas with contact interaction 
and periodic boundary conditions \cite{r6}. Later, 
the same authors have shown that the quantum transition properties 
of the attractive BEC in a 1D ring are strongly modified 
if the confining trap is rotating \cite{r7}. 
\par 
In this paper we investigate an attractive BEC in a 3D ring, 
taking into account transverse variations of the BEC width, 
showing that the phase diagram of the system reveals novel 
and peculiar structures. In particular, 
we prove that, contrary to the simple 1D case, 
the localized soliton has a limited 
existence and stability domain, which nevertheless 
strongly extends the stability domain of the uniform solution.  
Moreover, we find that the system supports also multi-peak 
solitons, which are energetically unstable but can be 
dynamically stable. Finally, we analyze the effect of 
a rotating ring.  In this case the multi-peak solitons 
are always energetically and dynamically unstable, while 
the one-peak soliton is stable 
in a domain that, for a fixed rotation frequency, critically 
depends on the system parameters. 

\section{Toroidal trap}

We consider a BEC confined in a toroidal potential 
given by 
\beq 
U(\rho ,\zeta ) = {1\over 2} m \omega_{\bot}^2 
\left( (\rho - R_0)^2 + \zeta^2 \right) \; , 
\eeq
where $\rho$ is the cylindric radial coordinate, 
$\zeta$ is the cylindric axial 
coordinate and $\theta$ is the azimuthal angle. 
The BEC has transverse harmonic confinement 
of frequency $\omega_{\bot}$ and the two characteristic 
lengths of the toroidal 
trap are $R_0$ and $a_{\bot}=(\hbar/(m\omega_{\bot}))^{1/2}$. 
In the remaining part of the paper 
we use scaled units: time in units of 
$\omega_{\bot}^{-1}$, length in units of $a_{\bot}$ 
and energy in units of $\hbar \omega_{\bot}$. 
To simplify the 3D Gross-Pitaevskii equation (GPE) 
we impose that the order parameter 
$\Psi (\rho ,\theta ,\zeta ,t)$ 
of the Bose condensate is the product of a generic azimuthal 
function $f(\theta ,t)$ 
and a radial Gaussian function, which has two 
variational parameters: the width $\sigma(\theta ,t)$ 
and the coordinate $R$ of the center of mass in the radial direction. 
The trial wave function is given by 
\beq 
\Psi (\rho ,\theta ,\zeta ,t) = 
{ f(\theta ,t) \over R^{1/2} }
{\exp{\left( -{(\rho - R)^2 + \zeta^2 \over 2 
\sigma(\theta ,t)^2}\right)} 
\over \pi^{1/2}\sigma(\theta ,t) } \; . 
\eeq 
We follow a procedure similar to that described in \cite{r8} 
inserting the trial wave function into the GPE Lagrangian 
density 
\beq 
{\cal L}
= \Psi^* \left[i {\partial \over \partial t} 
+ {1\over 2} \nabla^2 - U - {1\over 2} \Gamma |\Psi|^2 
\right] \Psi \; , 
\eeq 
where $\Gamma = 4\pi a_s N/a_{\bot}$, with $N$ the number
of condensed atoms and $a_s<0$ the attractive 
s-wave scattering length. 
The trapping potential of Eq. (1) has a cusp at the origin, 
a feature not usually present in experimental traps, 
but this is expected to have minor 
effects on physical properties if only a small fraction of particles 
are near the origin. For example, if $R/\sigma >3/2$, 
the fraction of particles belonging to the radial region 
$[0,R-\sigma ]$ is below $1\%$. 
In the range $1<R/\sigma <3/2$ the population near the origin 
is not so small and 
the results reported below are not fully reliable 
in this range. We integrate over $\rho$ and $\zeta$ 
coordinates \cite{r9}. 
In this way, from the Euler-Lagrange equations of the 
resulting effective Lagrangian density, we get the nonpolynomial 
Schr\"odinger equation (NPSE) 
\beq
\left[ i\frac{\partial}{\partial t} + 
\frac{1}{2}\frac{\partial^2}{\partial z^2} 
- T(n)\right] \psi = 0 \; , 
\label{dyn} 
\eeq
where $\psi(z,t)=f(\theta ,t)/R^{1/2}$ 
is the azimuthal 
wave function of the condensate 
with $z = R \theta$, $n=|\psi|^2$ is 
the density profile normalized to unity and 
$T(n)=\frac{dW(n)}{dn}$ with $W(n)=n(1-gn)^{1/2}$. 
The scaled interaction strength $g$ is given by 
$g = |\Gamma|/(2\pi) = 2N|a_s|/a_{\bot}$. 
In toroidal geometry the solution $\psi(z,t)$ must obey 
periodic boundary conditions $\psi(0,t)=\psi(L,t)$ 
with $L=2\pi R$. Within our variational approach 
the transverse width $\sigma$ 
of the BEC is given by 
\beq 
\sigma^2 = \left( 1 - g n \right)^{1/2} \; . 
\eeq 
For $g n \ll 1$ one has $\sigma \simeq 1$, $T(n)\simeq - gn +1$, 
and the NPSE reduces to the 1D GPE. In addition, we find 
that the variational parameter $R$ is implicitly given 
by the equation 
\beq 
R - {1\over R} \int_{0}^{2\pi R} 
\left|{\partial \psi\over \partial z}\right|^2 dz + 
{g\over 2 \sigma^2} \int_0^{2\pi R} |\psi|^4 dz = R_0 \; . 
\label{nuovo} 
\eeq
This formula shows that the effective radius $R$ of the BEC ring 
depends on the interaction strength $g$. 
We have verified that for a static and attractive 
BEC, the effective radius $R$ decreases very 
slowly by increasing $g$, 
while for a repulsive BEC the opposite is true. 
In practice, because for an attractive BEC the strength 
$g$ cannot exceed the scaled inverse density $n^{-1}$, 
the effective radius 
$R$ is always close to $R_0$: for a uniform BEC, 
where Eq. (\ref{nuovo}) becomes 
$R+g/\left( 4\pi R^2 (1-g/(2\pi R) )^{1/2} \right)=R_0$, 
it is easy to check that the relative difference between 
$R$ and $R_0$ is tipically of few percents and becomes 
$10\%$ only near the collapse. 
\par 
The very good accuracy of the NPSE in approximating the 3D GPE 
with a transverse harmonic potential has been verified in \cite{r8} 
for both positive and negative 
scattering length. In the derivation of NPSE one neglects 
the space and time derivatives of the width $\sigma(z,t)$. 
By including these terms one gets the coupled equations 
derived by Kamchatnov and Shchesnovich \cite{r10}, 
but it is not clear if these terms give an improvement. 
We have verified that, according to the 3D GPE, 
the single-peak bright soliton of an attractive BEC 
in an infinite cylinder collapses at the critical 
strength $g_c/2=0.676$ (see also \cite{r11}), a value 
very close to the NPSE prediction: $g_c/2=2/3=0.66{\bar 6}$ \cite{r8}. 

\section{Uniform and localized solutions}

The NPSE conserves both the norm of the wavefunction 
and the total energy $E$ of the configuration. 
The stationary solutions follow from Eq. (\ref{dyn}) by looking for
solutions of the form $\psi(z,t)=\phi(z) e^{-i\mu t}$ for some 
chemical potential $\mu$. 
The resulting non-linear eigenvalue equation 
is the static NPSE. 
In toroidal geometry, the uniform solution $\phi(z)=1/\sqrt{L}$ 
is always present for $g <L$, i.e. with density 
$N/L < a_{\bot}/(2|a_s|)$, and corresponds 
to the eigenvalue $\mu=T(1/L)$. 
In addition other, less trivial, profiles may be present. 
Beyond this limit (i.e. for $g>L$) the attraction is too 
strong and no regular solution is possible, leaving the 
BEC collapse as the only possibility. 
A generic (real) solution $\phi(z)$ of the stationary NPSE  
may be interpreted as the classical ``time" evolution of a fictitious particle 
moving in a potential $V(\phi)=\mu \phi^2 - W(\phi^2)$. As a consequence, 
the ``energy" conservation equation for this motion reads: 
\beq
\frac{1}{2}\left (\frac{d\phi}{dz}\right )^2 + V(\phi) = \epsilon  \; . 
\eeq
According to the values of the two parameters $\mu$ and $\epsilon$
two kinds of ``trajectories" may be realized: For $\mu > 0 $
and $\epsilon > 0$ the solution $\phi(z)$ oscillates between 
a positive and a negative value, thereby crossing zero, while
for $\epsilon < 0$ the solution $\phi(z)$ remains
always positive. In the first case the solutions are named 
{\sl nodal solitons} while in the second case {\sl nodeless 
solitons}. 

\begin{figure}
\centerline{\psfig{file=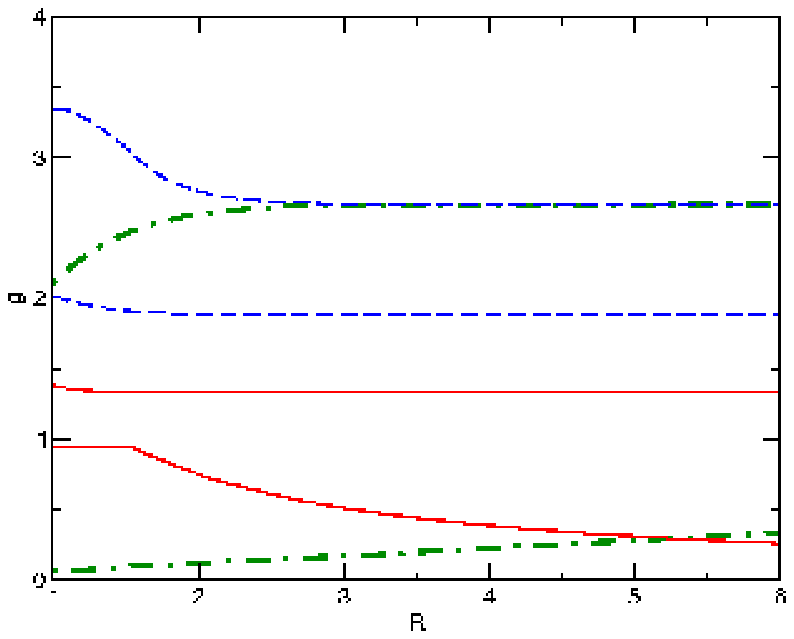,height=2.5in,clip=}}
{FIG. 1 (color online). Existence diagram in the $(R,g)$ plane, where
$g=2N|a_s|/a_{\bot}$ is the interaction coupling
and $R=L/(2\pi )$ is the azimuthal radius of the ring (in units $a_{\bot}$).
Uniform solution: exists for $g < 2\pi R$.
One-peak localized nodeless solution: exists between the
two solid lines. Nodeless two-peak localized solution
exists between the two dashed lines. The nodal two-peak
localized solution exists between the two dot-dashed lines.}
\end{figure}

\par 
For fixed $g$ and $L$, the two parameters $\mu$ and
$\epsilon$ are implicitly determined by the two consistency 
equations:
\beq 
L=\sqrt{2}N_s\,\int_{\phi_{min}}^{\phi_{max}} d\phi 
\frac{1}{\sqrt{\epsilon -V(\phi)}} 
\eeq
and 
\beq 
1=\sqrt{2}N_s\,\int_{\phi_{min}}^{\phi_{max}} d\phi 
\frac{\phi^2}{\sqrt{\epsilon -V(\phi)}} \; , 
\eeq  
where $N_s$ is the number of maxima of the soliton, 
i.e. the number of ``periods" of the corresponding orbit,
and $\phi_{min}$ ($\phi_{max}$) is the minimum (maximum)
value attained by $\phi(z)$ during the periodic oscillation.
As customary in Newtonian problems, 
the extremal values of $\phi$ are implicitly defined by the equation
$V(\phi)=\epsilon$ in terms of $\epsilon$ and $\mu$. 
The first equation comes from the commensurabilty requirement 
of the solitonic solution, which, after an integer number of 
periods must close without discontinuities. The second equation
is just the normalization condition of the wavefunction. 
\par
The two consistency equations (8) and (9) 
have been solved and the domain of 
existence of soliton solutions of given topology has been determined 
in the $(R,g)$ plane. Results are shown in Fig. 1. 
The numerical analysis shows that two solutions 
of same symmetry may be present in a portion of the existence domain. 
For sake of clarity in Fig. 1 we report the existence domain only 
for the one-peak soliton ($N_s=1$) and for the 
two-peak soliton ($N_s=2$). 
Note that the existence diagram of the 1D GPE is strongly 
different from that of the NPSE. In fact, as previously 
stressed, the 1D GPE does not take into account the transverse 
dinamics and, as a consequence, no collapse of 
solitonic solutions is predicted by 1D GPE. 

\section{Energetic stability}

We now turn to the discussion of the energetic stability \cite{r12} 
of the previously defined solitonic configurations. 
One can rigorously prove that the energetic stability 
can be expressed in terms of the eigenvalues $\lambda_l$ 
of $H\pm nT'$, where 
\beq 
H=-\frac{1}{2}\frac{d^2}{dz^2} +T(n)+n\frac{dT(n)}{dn}-\mu   
\eeq
and $n(z)$ is the density profile of the stationary solution. 
The stationary solution $\phi(z)$ is energetically stable 
only if either of the two conditions is satisfied: 
1) all the eigenvalues $\lambda_l$ are non-negative; 
2) a single negative eigenvalue $\lambda_0$ is present and 
$
\frac{dg}{d\mu}\le 0  
$.   
When we apply this general analysis to the simple case 
of the uniform stationary solution in toroidal geometry 
$\phi(z)=1/\sqrt{L}$ we find that this case satisfies the latter of 
the two conditions previously stated and the solution is stable 
until the second eigenvalue gets negative, triggering the 
instability. The resulting stability condition $(2\pi /L)^2 + 4nT' \ge 0$ 
explicitly becomes: 
\beq 
{\pi^2 \over gL} \left ( 1-\frac{g}{L}\right )^{3/2} \ge 
\left (1-\frac{3g}{4L}\right ) \; . 
\eeq
This formula reduces to $\pi^2/(gL)$ for large $L$ (1D limit), 
that is precisely the result one finds with the 1D GPE \cite{r6}.  
The stability analysis of the solitonic configurations, however,
does not allow for a general analytic solution and the 
eigenvalue equations of $H\pm nT'$ must be investigated numerically. 

\begin{figure}
\centerline{\psfig{file=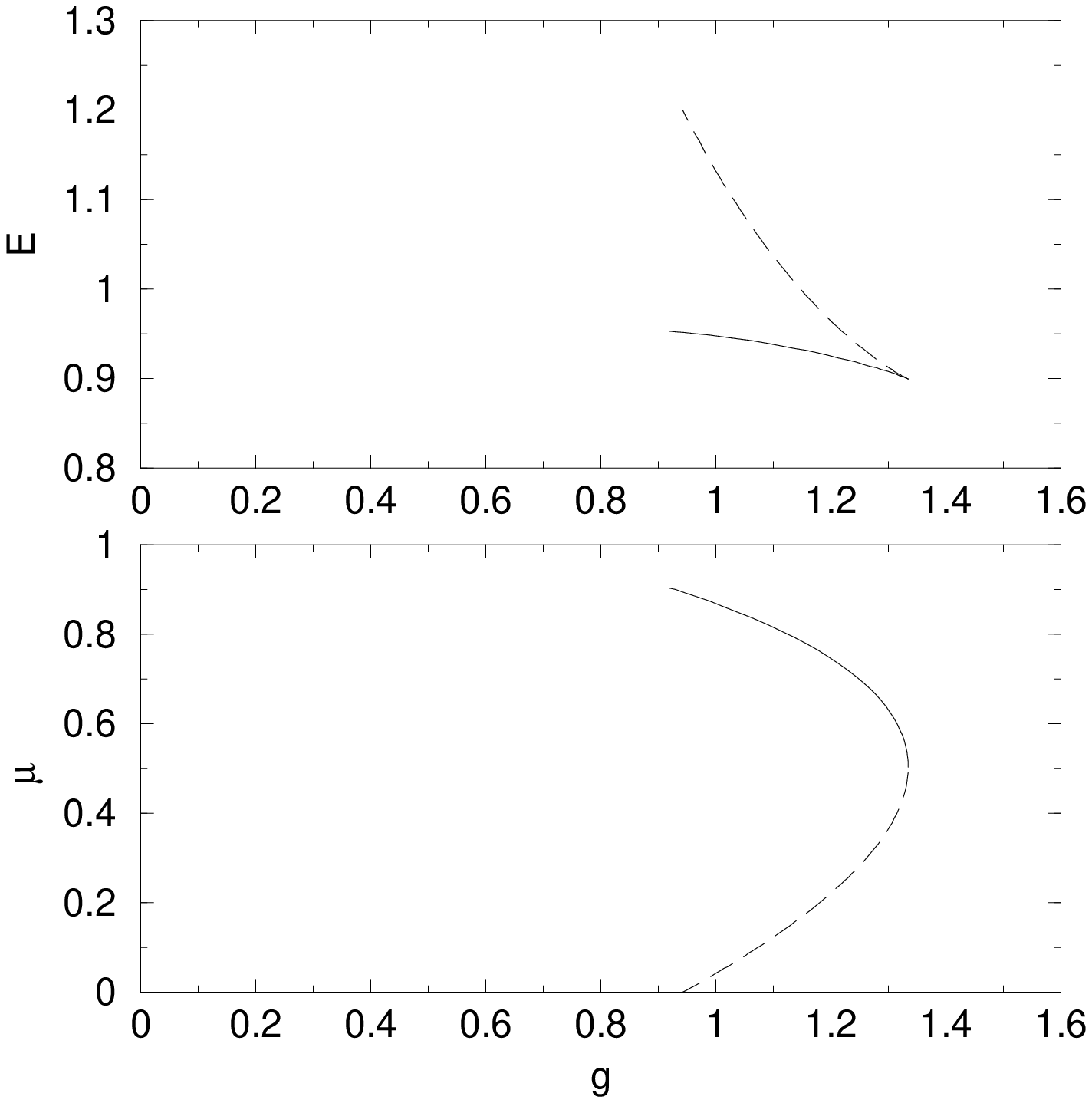,height=3.5in,clip=}}
{FIG. 2. Energy $E$ and chemical potential $\mu$ of the one-peak solitons
as a function of the coupling $g$ for $L=10$.
The energetically stable solutions are the solid ones,
the unstable solutions are the dashed ones.}
\end{figure}

\par 
The operators $H\pm nT'$ may be numerically diagonalized
by introducing a finite mesh in the interval $0 \le z < L$ and 
approximating the differential operator with the corresponding
finite difference operator. 
The two resulting equations then give rise to 
a matrix eigenvalue problem. The numerical results show that: 
\begin{itemize}
\item 
in the regions where the uniform solution satisfies the energetic 
stability condition $\frac{dg}{d\mu}\le 0$ no 
other solitonic wavefunction can be stabilized; 
\item  
only the one peak soliton is energetically stable in a portion of
the domain where it is defined; 
\item 
when distinct one peak solutions exist for 
the same values of $g$ and $L$, 
the soliton is stable only in the branch corresponding 
to the lowest energy. 
\end{itemize}
The latter remark is illustrated in the upper panel of 
Fig.2 where the energy $E$ of the one-peak solution 
is shown as a function of $g$ for $L=10$. 
It is also interesting to plot the 
chemical potential $\mu$ versus $g$ 
in the stable and unstable branch. The case of $L=10$ previously analyzed 
is shown in the lower panel of Fig. 2, 
where it appears that the onset of instability 
corresponds to an extremum of the coupling constant $g$ as a 
function of the chemical potential. In fact, this immediately follows 
from the 
previous analysis which led to $\frac{dg}{d\mu}\le 0$: 
if $H+nT'$ admits a single negative 
eigenvalue, $\frac{dg}{d\mu}=0$ signals the onset of the instability.

\begin{figure}
\centerline{\psfig{file=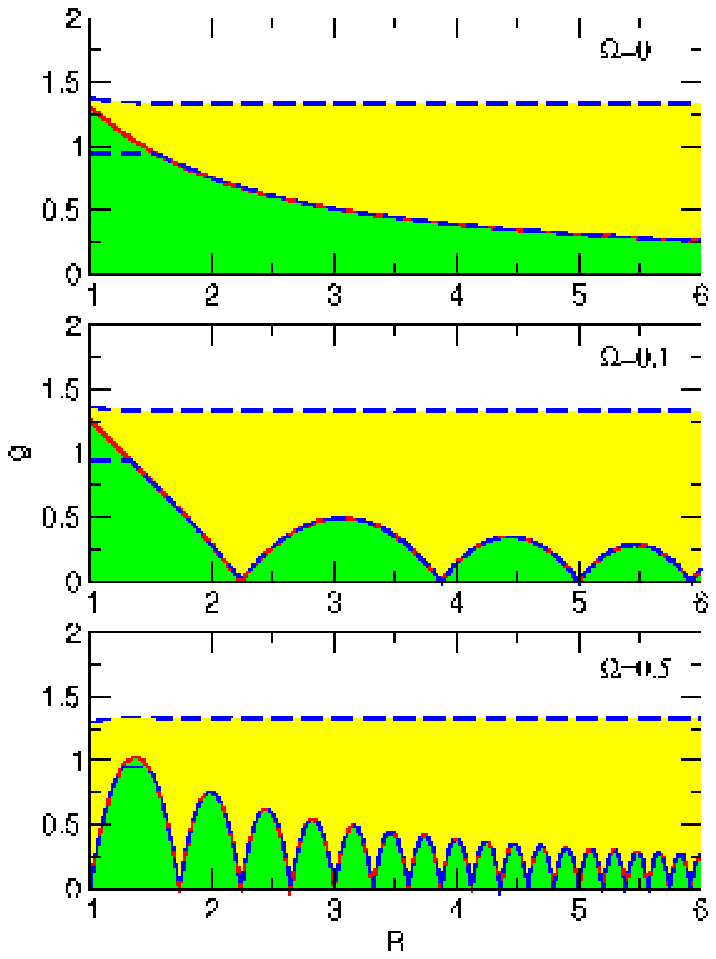,height=3.9in,clip=}}
{FIG. 3 (color online). Attractive BEC in a
ring rotating with frequency $\Omega$.
Energetic-stability diagram in the plane $(R,g)$.
The uniform solution is the ground-state only below the solid line.
The one-peak localized solution exists between the
two dashed lines but it is the ground-state only between
the solid line and the upper dashed line.
Note that for almost all $R=L/(2\pi )$ the solid curve
is superimposed to the lowest dashed curve: 
only for $1<R<1.5$ the lowest dashed line is below 
the solid line. }
\end{figure}

\par 
The energetic stability region of stationary solutions of the NPSE 
in a ring are shown in the top panel ($\Omega =0$) 
of Fig. 3. Below the solid curve the uniform solution 
is energetically stable. 
The one-peak soliton, which exists between the two dashed curves, 
is energetically stable in the domain limited by the solid 
and the upper dashed line. 
In the remaining regions of the phase diagrams 
no energetically stable stationary solution is present 
and the BEC is expected to collapse. 
Note that for large $R=L/(2\pi )$ the upper dashed 
line tends to $g = 4/3$, 
that is the formula one finds for the collapse of a bright soliton 
in a infinite cylinder (see L. Salasnich 
{\it et al.} 2002 in \cite{r3}). 
It is not difficult to show that with a large $R$ 
the existence domain  of a $N_s$-peak bright 
soliton is instead given by 
\beq
0< g < {4\over 3} N_s  \; . 
\eeq  
See for instance the existence domain of the two-peak solitons shown 
in Fig. 1. As previously stressed upper bounds 
do not exist within the 1D GPE approach: the collapse 
of single and multiple bright solitons is due to the 
transverse dynamics of the condensate. 
\par 
In the other panels of Fig. 3 
it is shown the effect of a rotating toroidal trap 
on the attractive BEC. The analysis is developed by observing that 
one has to include the centrifugal operator 
\beq 
- \Omega {\hat M_{\theta}} = 
i \Omega {L\over 2\pi } {\partial \over \partial z}
\eeq 
into Eq. (\ref{dyn}), where $\Omega$ is the rotation 
frequency (in units of the frequency $\omega_{\bot}$ of 
the harmonic transverse confinement) 
and ${\hat M_{\theta}}$ is the azimuthal angular momentum. 
As shown in Ref. \cite{r7} by using the 1D GPE, the uniform state 
of the attractive BEC is superfluid, i.e. 
it exists a critical frequency $\Omega_c$ 
below which the uniform state remains stationary, and only above 
this critical frequency the uniform state rotates. 
In general the stationary uniform solution is thus 
given by $\psi(z) =e^{i 2\pi z j /L}/\sqrt{L}$, 
where the integer number $j$ is a function of $\Omega$ and $L$, namely 
\beq 
j(\Omega, L) =int[{\Omega L^2 \over 4\pi^2} 
+ {1\over 2}]  \; , 
\eeq 
with $int[x]$ the maximum integer that does not exceed $x$. 
The localized soliton solution has a different behavior: 
its angular momentum is not quantized \cite{r7} and 
this means that the quantum phase 
transition from the uniform to the localized state suppresses 
the superfluidity of the system \cite{r7}. 
Setting $\psi(z) = \phi(z) e^{i\alpha(z)}$, where both 
$\phi (z)$ and $\alpha(z)$ are real, 
from the stationary NPSE with the centrifugal operator of 
Eq. (13) one finds 
\beq 
{d\alpha \over dz} = \Omega {L\over 2 \pi} + {c \over \phi^2} \; , 
\eeq
where the constant $c$ is given by the equation 
\beq 
\Omega R^2 + {c \over 2\pi } 
\int_0^L {dz\over \phi(z)^2 } = j \; . 
\eeq
In addition, the function $\phi(z)$ is obtained 
from the two consistency equations (8) and (9) 
with $V(\phi )$ now given by 
\beq 
V(\phi ) = \mu + {1\over 2} 
\left({\Omega L\over 2\pi}\right)^2
- W(\phi ) + {c^2\over 2 \phi^2} \; , 
\eeq 
where $W(\phi ) = \phi^2 (1 - g \phi^2)^{1/2}$. 
\par 
Following the previous analysis one finds 
that the energetic stability condition reads 
\beq 
1 - 4 \left( {\Omega L^2\over 4\pi^2} - j(\Omega, L)  \right)^2 
\le  
{g L\over \pi^2} {(1-\frac{3g}{4L}) 
\over ( 1-\frac{g}{L} )^{3/2} }    \; . 
\eeq  
In the 1D limit of large $L$ the previous formula gives 
$1 - 4 \left( {\Omega L^2\over 4\pi^2} - j(\Omega, L)  \right)^2 \ge 
gL/\pi^2$, that is the results found in Ref. \cite{r7}. 
Fig. 3 shows the energetic stability diagram 
in the plane $(R,g)$ for two non-zero values 
of the rotating frequency $\Omega$. Below the solid line 
the stability condition holds and the 
uniform state is energetically stable. The periodic structure 
for $\Omega\neq 0$ is a consequence of the periodic 
quantization of the angular momentum $j(\Omega, L)$. 
In particular, for a fixed $\Omega$, 
the solid line touches the horizontal axis $g=0$ 
for discrete values of $R=L/(2\pi )$, 
which correspond to jumps in the quantum number $j$. 

\begin{figure}
\centerline{\psfig{file=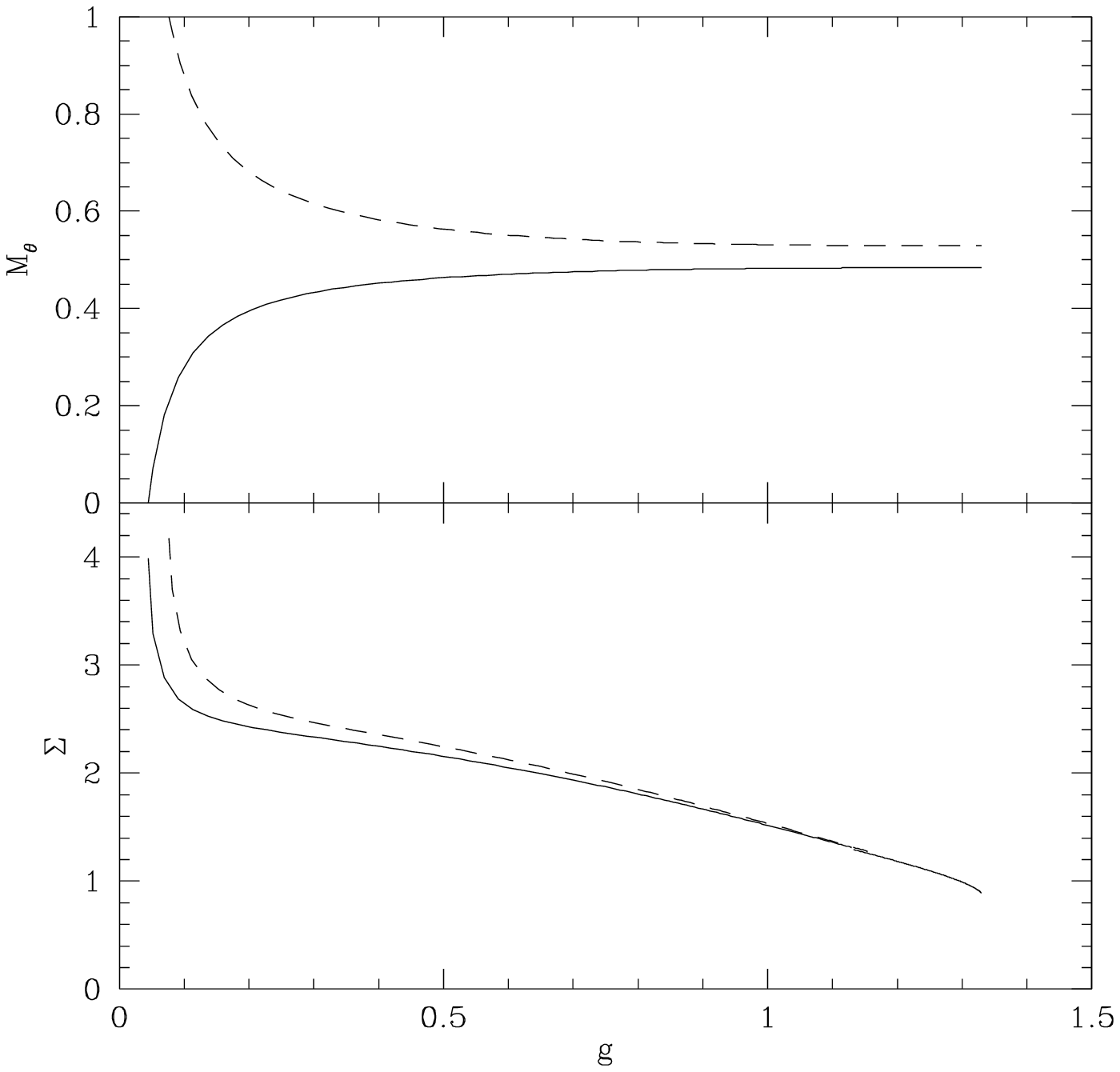,width=3.4in,height=3.6in,clip=}}
{FIG. 4 Angular momentum $M_{\theta}=\langle \hat{M}_{\theta} 
\rangle$ 
and azimuthal width $\Sigma = \langle z^2 \rangle^{1/2}$ 
of the one-peak soliton 
as a function of the coupling $g$ for $\Omega = 0.1$ 
and two values of $R$: $R=2.2$ solid line, 
$R=2.3$ dashed line.}
\end{figure}

The one-peak solitonic solution 
exists between the two dashed lines and it is energetically 
stable between the solid line and the upper dashed line. 
Interestingly, Fig. 3 shows that for large $R$ the 
lower dashed line and 
the solid line practically coincide. 
As in the non rotating case, we find that also for  
$\Omega\neq 0$ solutions with more than one peak are not 
energetically stable. 
\par 
In Fig. 4 we plot the angular momentum $M_{\theta}$ and 
the azimuthal width $\Sigma$ of the rotating 
one-peak bright soliton. The figure shows that the angular 
momentum is not quantized: it approaches the 
value $M_{\theta}=\Omega R^2$ of a ``classical particle''  
for $g$ close to the collapse ($g\simeq 4/3$), 
but it becomes quantized, 
i.e. $M_{\theta}=j$, where $j$ depends on $\Omega$ and $R$, 
for the small value of $g$ that gives the transition 
to the uniform solution. Obviously, when the angular 
momentum becomes quantized the width $\Sigma$ of the 
bright soliton coincides with that of the uniform solution. 
Fig. 4 also shows that the width $\Sigma$ is independent 
on the ring radius $R$ as the bright soliton is 
close to the collapse; in this case $g\simeq 4/3$ and 
$\Sigma \simeq 0.85$. 

\section{Dynamical stability}

It is important to stress that energetic stability 
implies dynamical stability but the converse 
is not true \cite{r12}. In order to investigate 
the dynamical stability of stationary solutions 
in the ring one can solve the Bogoliubov-de Gennes (BdG) 
equations which give the elementary excitations 
$\epsilon_l$ ($l=1,2,3,...$) of the system: the appearence of a complex 
excitation signals dynamical instability while a negative 
excitation implies energetic instability \cite{r12}. 
For the uniform solution $\psi(z) =e^{i 2\pi z j /L}/\sqrt{L}$ 
one finds 
$$
\epsilon_l = 
-\left( \Omega -{4\pi^2\over L^2} j(\Omega,L) \right) l 
$$
\beq 
+ {1\over 2} 
\left\{ ({2 \pi l \over L})^2 
\left[ ({2 \pi l \over L})^2 - 
{4g\over L} {(1 + {3g\over 4L})\over (1-{g\over L})^{3/2} } 
\right] 
\right\}^{1/2} \; . 
\eeq 
This result confirms that the uniform 
solution is energetically stable if the previously written 
stability condition is satisfied; 
moreover it shows that the dynamical stability 
of the uniform solution is independent on $\Omega$. 
For localized solutions the BdG equations are computationally 
rather demanding. For this reason we have analyzed the dynamical 
stability by numerically solving  
the time-dependent NPSE taking as initial condition 
the stationary localized solution $\phi(z)$ with a very weak 
perturbation. 

\begin{figure}
\centerline{\psfig{file=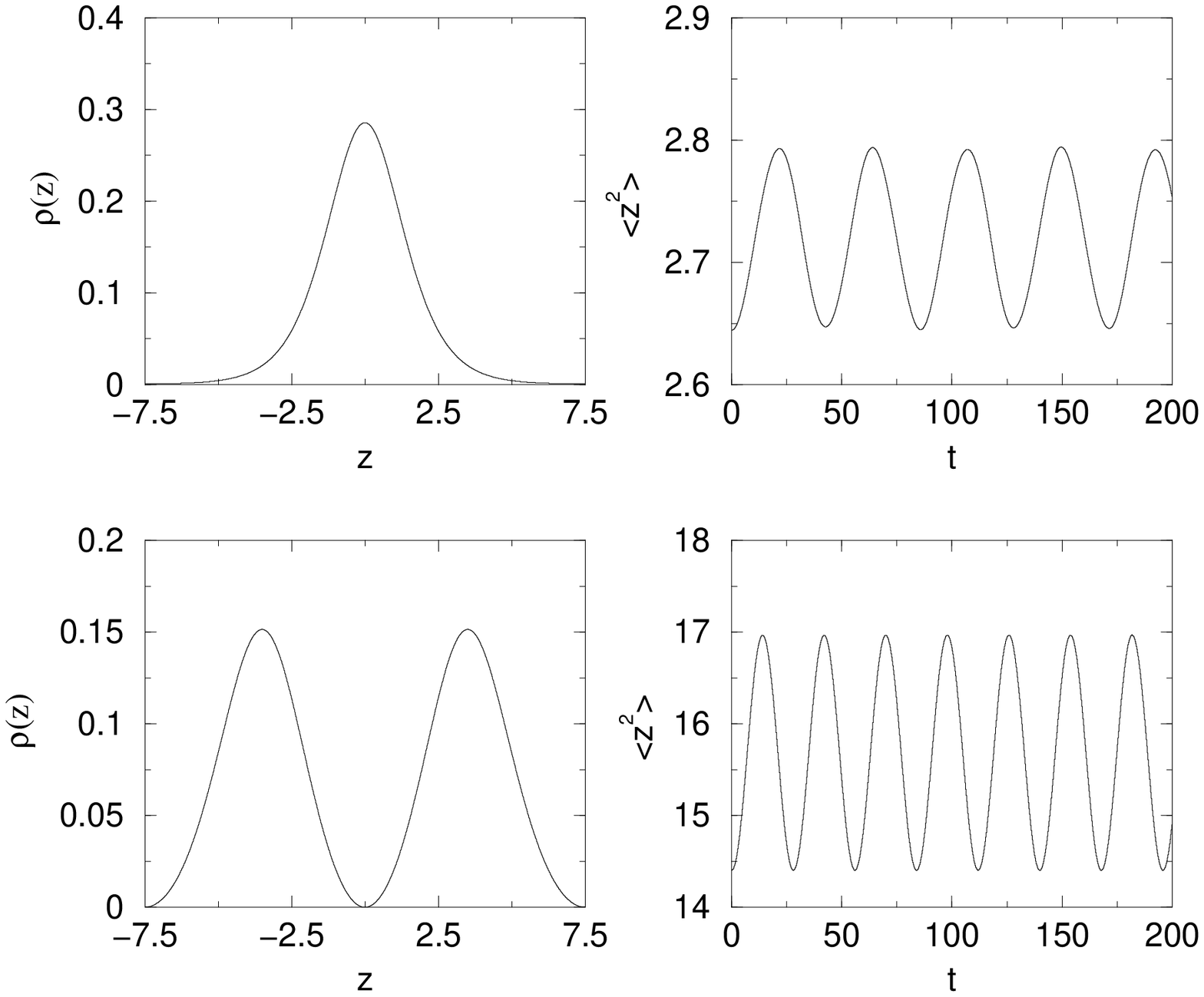,width=3.2in,height=3.2in,clip=}}
{FIG. 5. Left panels: density profile $\rho(z)$
of the solitonic solutions. Right panels: time-dependence
of the mean squared widths $\langle z^2 \rangle
- \langle z \rangle^2$
for the weakly perturbed solitonic solutions.
Ring axial lenght: $L=15$. Interaction strength: $g=1$.
Rotational frequency: $\Omega=0$. }
\end{figure}

\par
In Fig. 5 we plot the density profile 
$\rho(z)=|\phi(z)|^2$ of the one-peak and the nodal 
two-peak solutions, choosing $L=15$, $g=1$ and two 
values of $\Omega$. 
In addition, we plot the time-evolution of the mean squared  
width $\langle z^2\rangle - \langle z \rangle^2$. 
Its behavior reveals that 
these solitonic solutions are dynamically stable. 
We have investigated the dynamical stability for various 
initial conditions. For $\Omega =0$ we have found that: 
\begin{itemize}
\item 
the one-peak soliton is dynamically 
stable where it exists; 
\item 
the nodal two-peak soliton is dynamically 
stable in the plane $(R,g)$ only below the upper 
curve of existence of the one-peak soliton; 
\item 
the nodeless two-peak soliton is dynamically unstable.  
\end{itemize} 
Similar results are found for solitonic 
solutions with a larger number $N_s$ of peaks. 
\par 
The case $\Omega\ne 0$ leads to similar results, keeping in mind, however,
that nodal solitons do not exist under rotation. 
For high rotational frequencies, namely when $\Omega$ approaches 
the frequency of transverse harmonic confinement (that is $1$ in our 
units), the effect of the centrifugal force on the transverse 
dynamics becomes relevant. 
For a given angular momentum $j\simeq R^2 \Omega$, 
the centrifugal force increases the effective radius $R$ 
of the BEC ring. For a rotating and uniform ideal BEC 
one easily finds from Eq. (\ref{nuovo}) that $R=R_0/(1-\Omega^2)$. 
This formula holds also for an azimuthally localized ideal BEC. 
At the critical frequency $\Omega =1$ 
the radius $R$ diverges and this means that the Bose condensate 
is no more confined. As in the non-rotating case, an investigation 
of Eq. (\ref{nuovo}) shows the effect of the interaction strength $g$ 
on the effective radius $R$ is negligible 
for an attractive BEC. 

\section{Conclusions}

We have predicted novel quantum phases 
for an attractive Bose condensate in a ring. Our results 
can be experimentally tested with the optical and magnetic 
traps recently developed \cite{r5,r13}, where a time-dependent 
stirring potential can be used to set into rotation 
the system. For instance, by choosing $10^{3}$ $^7$Li atoms 
($a_s=-1.4$ nm) in a toroidal trap with $L\simeq 25$ $\mu$m and 
$a_{\bot}\simeq 3$ $\mu$m, a sequence of transitions between the 
uniform state and solitonic configurations takes place 
when the stirring frequency $\Omega$ is varied 
between $0$ and $1$ kHz. 
These experiments will open the way to the observation of 
amazing quantum phenomena like the solitonic condensate 
without superfluidity and the dynamically induced phase 
transition from uniform to localized states. 

\section*{acknowledgements}

The authors thank F. Dalfovo, A. Recati, S. Stringari and 
C. Tozzo for useful suggestions.

\end{document}